# Implementation and Complexity of the Watershed-from-Markers Algorithm Computed as a Minimal Cost Forest

Petr Felkel

VrVis Center, Lothringerstraße. 16/4, A-1030 Vienna, Austria, *www.vrvis.at*, contact address: *felkel@vrvis.at*

Mario Bruckschwaiger and Rainer Wegenkittl

TIANI Medgraph GmbH, Campus 21, Liebermannstraße A01 304, A-2345 Brunn am Gebirge, Austria, *www.tiani.com*

**Abstract**
*The watershed algorithm belongs to classical algorithms in mathematical morphology. Lotufo et al.[1] published a principle of the watershed computation by means of an Image Foresting Transform (IFT), which computes a shortest path forest from given markers. The algorithm itself was described for a 2D case (image) without a detailed discussion of its computation and memory demands for real datasets.*

*As IFT cleverly solves the problem of plateaus and as it gives precise results when thin objects have to be segmented, it is obvious to use this algorithm for 3D datasets taking in mind the minimizing of a higher memory consumption for the 3D case without loosing low asymptotical time complexity of $O(m+C)$ (and also the real computation speed). The main goal of this paper is an implementation of the IFT algorithm with a priority queue with buckets and careful tuning of this implementation to reach as minimal memory consumption as possible.*

*The paper presents five possible modifications and methods of implementation of the IFT algorithm. All presented implementations keep the time complexity of the standard priority queue with buckets but the best one minimizes the costly memory allocation and needs only 19–45% of memory for typical 3D medical imaging datasets.*

*Memory saving was reached by an IFT algorithm simplification, which stores more elements in temporary structures but these elements are simpler and thus need less memory.*

*The best presented modification allows segmentation of large 3D medical datasets (up to $512 \times 512 \times 680$ voxels) with 12- or 16-bits per voxel on currently available PC based workstations.*

## 1. Introduction

The idea behind application of watersheds (WS) in a grey-scale image segmentation is very simple. The aim is to find objects and their borders. As the object borders are loci of the highest gradient in the image, the gradient image is taken as an input height field (topographical image) that is step by step immersed into water. During the process of immersion, water forms *catchment basins*, which correspond in an ideal case to the objects. The creation of catchment basins begins in local minima of the gradient image. When these catchment basins meet (touch), a dam is built to prevent them to join. These dams then form the *watershed lines*, which should correspond to object boundaries.

The watershed algorithms starting in local minima suffer from oversegmentation if applied to images with noise, as mentioned, e.g., by Sonka[2]. For a region growing it is better to use a *watershed from markers*, where the creation of catchment basins begins in these markers and only a given number of objects are found.

A very interesting algorithm for computation of watersheds from given markers by means of the *Image Foresting transform* (IFT) with a priority queue has been proposed by Lotufo and Falcao[1] but without detailed implementation issues. An advantage of this algorithm is its higher resolution, as it can correctly segment blobs connected by an one-pixel thick region.

The tuning of the implementation of the priority queue via buckets to reach as minimal memory consumption as possible, keeping the speed of the standard priority queue with buckets, is the aim of this paper. The best of five presented algorithm variants needs only about one third of memory in a typical case of 3D medical imaging datasets.

The paper has the following structure. Section 2 describes basic terms, Section 3 the IFT algorithm as proposed by Lotufo and Falcao[1], Section 4 describes five modifications of the IFT algorithm, Section 5 the datasets and the worst and average case analysis of the algorithms. Conclusions and the future work follow in Section 6.

## 2. Definitions and notations

Let D(x,y,z) be a volumetric dataset of size $x \times y \times z$ voxels. Let this dataset be taken as an undirected graph $G = (V,E)$ with node set $V$ of size $n = xyz$ and arc set $E$ of size $m = xy(z-1) + x(y-1)z + (x-1)yz$, defined by a 6-connectivity of neighboring voxels.

Let $w(p,q)$ be a weight of the arc $(p,q)$ between the neighboring voxels and $f(u)$ a value stored in voxel (node) $u$. The nonnegative integer arc weight is for the watershed algorithm defined as $w(p,q) = |f(p) - f(q)|$. Note that this formulation of the arc weight gives a very simple approximation of a complete 3D gradient as known in common sense but works correctly within this algorithm. Let $C$ be a maximal arc weight in the dataset.

The *shortest-path forest problem* finds for each graph node the shortest path connecting it to its nearest root node. In case of the *watersheds from markers*, the root nodes are defined by the user as markers of two types: the IN marker for the objects and the OUT marker for the background. The path length is measured by the *path cost* defined as follows:

Let *path cost* $C(p,q) = [M,d]$ in the path from node $p$ to node $q$ be defined as a pair of values in lexicographic order. The first component (with a higher priority) is the *maximum arc weight M*. The second component $d$ is a distance from the end of the path to the nearest node with a lower path cost. More precisely, for the path $(p,q) = (v_1, v_2, \ldots, v_n)$, where $v_1 = p$, $v_n = q$ and $v_i \neq v_j$ for $i \neq j$, the path cost components are defined as follows: $M = M(n) = \bigvee_{i=1}^{n-1} w(v_i, v_{i+1})$ and $d = \min\{j : M(i) = M(i-j), j = 0, 1, \ldots, i-1\}$. According to Lotufo and Falcao[1], the second component $d$ reflects the flooding process when the water reaches a plateau in the relief.

The *shortest-path cost* between two nodes $C^*(p,q)$ is defined as the smallest lexicographic cost of all paths between $p$ and $q$. The detection of the shortest-path cost from all nodes results in the detection of the "nearest" marker for each node, i.e., labeling of the voxel as being part of the object or of the background.

Lotufo and Falcao[1] describe the physical meaning in the frame of a topological relief of the image as follows: "The arc weight is the height of the wall between nodes and the shortest-path cost is the minimal height where the water coming from two points merges. ..., the second component of the lexicographic cost allows the partition to be at the medial line of the plateau."

The order of en-queueing of the markers into the priority queue is important, as it determines the order of processing of the markers and also the shape and the size of the resulting labeled area. The markers defined in first place acquire areas that are a little larger (of about one-pixel on plateaus).

The theoretical most efficient known algorithm for the shortest-path forest problem is the Dijkstra's algorithm[3], which can be efficiently implemented by means of a priority queue. It runs in $O(m)$ time plus the time required to perform queue operations.

If the arc weights are real numbers and only binary comparisons are used in the heap implementation the best possible time for the Dijkstra's algorithm is $O(m + n\log n)$, as described e.g. by Ahuja et al.[4] If all the arc weights are integers of moderate size $C$ (which is true in the case of medical datasets) special algorithms exist that differ in the implementation of the priority queue, running from a typical time of $O(m + nC)$ (the Dial's[5] implementation with buckets) to the best known time of $O(m + n\sqrt{\log C})$ (Two-Level Radix heaps by Ahuja et al.[4], which combine Radix heaps and Fibonacci heaps). These values are valid for path cost defined as a sum of the weights along the path. The definition used for the watershed algorithm as described above in this section results in a time complexity of $O(m + C)$ (Details in Section 5.2).

The memory complexities of these algorithms are: $O(n + C)$ for the Dijkstra's[3] and the Dial's[5] algorithms (up to $O(n)$ if implemented with hashing but the time is then not deterministic), and $O(m + \log C)$ for One level Radix heaps by Ahuja et al.[4]. For Two level Radix heaps the memory consumption was by Ahuja et al.[4] not mentioned.

## 3. IFT algorithm by Lotufo and Falcao

Lotufo and Falcao[1] implemented the watershed algorithm as an Image Foresting Transform (IFT). They accent the necessity of a priority queue with a FIFO restriction for the correct handling of plateaus. In the algorithm, $C(p)$ is the cost path from $p$ to its nearest marker, $L(p)$ is the input marker image and also the result of the watershed partitioning, $flag(p)$ differentiates finally labeled nodes from non-processed or temporarily labeled nodes. The IFT algorithm works as follows:

```
1. Initialization
   a) for all nodes p do
         flag(p) = TEMP;
   b) for all non-marker nodes p do
         C(p) = infinity;
   c) for all marker nodes p do
         C(p) = 0; EnQueue(p,0);
2. Propagation
   while QueueNotEmpty() do
      a) v = DeQueueMin;
      b) flag(v) = DONE;
      c) for each p neighbor of v with
              flag(p) == TEMP do
         if max{C(v), w(v,p)} < C(p) then
            A) C(p) = max{ C(v), w(v,p) };
               L(p)=L(v);
            B) if IsInQueue(p) then
                  DeQueue(p);
            C) EnQueue(p, C(p));
```

The algorithm partitions all the graph nodes into two sets: permanently (DONE) and temporarily (TEMP) labeled. At each iteration step it selects a temporarily labeled node with a minimum cost from the priority queue as the next node to be scanned. Once a node is processed it becomes permanently labeled. The algorithm terminates when all nodes become permanently labeled, i.e., when the queue is empty. As the path cost is a non-decreasing function, the important property of the algorithm is that when the node becomes permanent its path cost is the final optimal path cost. Another important property is that the second component $d$ of the path cost does not need to be stored, as the FIFO priority queue keeps the second component intrinsically sorted.

In the initialization phase, all the nodes are labeled as temporary (line 1a), the markers have assigned a cost to 0 (line 1c), all other nodes to infinity (line 1b). The markers are entered into the priority queue. The queue insertion sequence for markers of the same type (e.g., only background ones—OUT) is not important. On the other hand, the order of insertions of the markers of a different type into the queue is important, because the following processing is also done in the FIFO order and the first come marker "wins" a larger region. Therefore two separate input lists of object and background markers are used and the algorithm starts width en-queueing of the object markers.

During the propagation step, the node $v$ with a minimal path cost is removed from the queue and marked as permanent (lines 2a,b). All its temporarily labeled neighbors $p$ are then tested and if the path cost through the permanent node $v$ is smaller than the temporary cost associated with node $p$ (line 2cA) the temporary cost and label are updated. If the node was already in the queue it is removed (line 2cB) and finally a node is enqueued with the priority of a new path cost (line 2cC).

For a better understanding of the role of the distance component $d$ it will be discussed in a greater detail: The algorithm processes the nodes from the lowest path cost to the highest, as the flooding by water proceeds. That also means while processing a given level of path cost, all nodes with a lower path cost are already processed! The same is also true for the distance component $d$, as all nodes with a lower distance are already processed and only the nodes with distance $d$ and $d+1$ exist in the queue. More precisely, the nodes with distance $d > 0$ exist only for the current path cost level $C(v)$, which means, for the queue implemented with buckets, in the current bucket.

If the processed node $v$ is on the given "water" level, which means it has the path cost equal to $C(v)$, there are three possibilities for the neighboring node $p$:

1. The path cost $C(p)$ *will not be updated*, as the path through the node $v$ would be more expensive. The same is true for $d$, which survives unchanged (test in line 2c is false),
2. the path cost *will be updated* to max$\{C(v), w(p,v)\}$, as it is cheaper to go through the node $v$ (test in line 2c is true):

   a) *Path cost increases* (the arc weight $w(p,v) > C(v)$), $C(p) = w(p,v)$ and a new distance $d(p) = 0$,
   b) *path cost value remains, plateau* ($w(v,p) \leq C(p)$), $C(p)$ remains unchanged and a new distance $d(p) = d(v)+1$ is computed.

In both cases where the priority queue is updated the FIFO restriction handles correctly the insertion of a new node in the end of all already stored nodes. In the case 2a), a new node is inserted as the last node with the path cost equal to $C(p)$. All stored nodes with the cost higher than $C(p)$ have the distance $d = 0$.

In the case 2b), which handles plateaus, neighboring nodes are "reinserted" with the cost $C(v)$ unchanged but with the distance $d$ incremented ($d(p) = d(v)+1$). The nodes already stored in the queue with this path cost have associated two possible values of the distance $d$: The distance is either equal to $d(v)$ (the node is waiting for processing at this level of $C(v)$ and $d(v)$) or the distance is equal to $d(v)+1$ (the neighboring node that has been processed at the current level of $C(v)$ and $d$). Newly processed nodes are successively inserted after all nodes with values $C(v)$ and $d(v)+1$.

If an implementation with buckets is used both the cases 2a) and 2b) represent inserting of the nodes in the end of the appropriate bucket. The principle of buckets is explained in Section 4.

## 4. Implementation of the IFT algorithm

Falcao and Lotufo[1] did not discuss the implementation of the algorithm in detail. They only noted the application of buckets for implementation of the priority queue similar to Dial[5]. Buckets can be used, as the maximal arc weight $C = w_{max}(p,q)$ for a given input grey-scale volume is known—it is the maximal absolute difference of voxel values ($C = max|f(p) - f(q)|$ for all neighboring voxels). $C$ is also called *MaxDiff* in the following text and it is upper-bounded by the maximal value store-able in the dataset. This maximal store-able value is in this paper called a dataset *precision* ($C_p$) and it is given by the 12–16 bit *representation of the numbers* in the dataset.

They also proposed to spare the line 2cB in the IFT algorithm, which simplifies the data structure for priority queue implementation but results in more elements stored in the priority queue, as multiple entries for one voxel will appear in the queue. Possible methods for the implementation of the IFT algorithm will be discussed below in this section.

The IFT algorithm of Falcao and Lotufo[1] follows the ideas proposed by Dial[5] and uses *buckets* for the priority queue implementation to overcome the logarithmic time complexity of heap based approaches (for handling of temporary labeled nodes), which is identified as a bottleneck of the whole algorithm. The bucket $k$ stores all temporarily labeled nodes whose distance labels are equal to $k$ (or fall within a certain range, as in other approaches). The nodes in buckets can be stored in double-linked lists of nodes, to which the array of pointers with a size equal to the number of buckets $C$ points.

The Dial's implementation requires $nC + 1$ buckets in the worst case, where $C$ is the maximal cost. However, Ahuja et al.[6] proved that for a graph with a maximum cost step $C$ (network with a maximal arc length $C$) only $C + 1$ buckets are necessary to maintain all the temporarily stored labels in a circular queue. As the algorithm uses the maximum on the path as a path cost, even a circular queue is not needed and the number of buckets $C$ is upper-bounded by the precision $C_p$ of the dataset.

In the following text five variants of the algorithm implementation will be discussed and results of measurements of their time and memory consumption if applied on real 3D medical datasets will be shown. The different modifications of the algorithm use different levels of simplification of the data structures necessary for the implementation and therefore also change the memory complexity of the algorithm in both average and worst cases. The algorithms have been tested in order to know which of the implementations has the minimal average memory consumption keeping also the time complexity of $O(m+C)$. The tested algorithm modifications were:

I) A complete IFT algorithm with dequeuing and a fixed volume for queue (Section 4.2),
II) a complete IFT algorithm with dequeuing and a dynamical queue (Section 4.3),
III) an IFT algorithm without dequeuing of the entries with a higher path cost from the queue (Section 4.4),
IV) a modification without dequeuing that stores all the costs in the queue (Section 4.5), and finally
V) an improvement of the modification (IV) which saves the space necessary for the queue implementation (Section 4.6).

In the following subsections all five variants of the algorithm modifications will be explained.

### 4.1. Common and algorithm specific data structures

We have designed the data structures in all algorithm modifications with the aim of achieving a constant time complexity for the queue modification. The constant time complexity is valid for all queue operations except of searching for the first non-empty bucket in the operation *QueueNotEmpty()*, which has a complexity of $O(C)$.

The following data structures are common to all presented implementations:

- *Input dataset*—volume of scalar values of size $n = x \times y \times z$, stored with precision $C_p = 4096$ or $65536$, given by the 12- to 16-bit number representation. The dataset has $m$ arcs (as described in Section 2),
- two *fixed arrays of pointers* to the beginnings and ends of the buckets. The arrays have indices in range $\langle 0, C \rangle$ where $C = \text{MaxDiff}$ is the maximal absolute difference of neighboring values in the dataset. As the upper bound for this number is given by the precision $C_p$ of the the dataset, $\text{MaxDiff} = O(C_p)$ and therefore $C_p$ can be directly used as the higher index in the following sections if the maximal value $C$ is not available,
- *temporary volume of flags*—one bit value that distinguishes TEMP and DONE status of each node in the volume. The flags are stored either in a separate bit-volume or together with the position as one bit from the 4-byte value,
- *resulting volume of labels*— one bit for each label, as only the object and the background are distinguished.

The implementation specific data structures, which differ in the modifications of IFT, are as follows:

- *Buckets of elements*—implemented as a fixed volume or dynamically as double-linked lists, single-linked lists or lists stored in structures called bricks, which will be described in Section 4.6,
- *temporary node costs*—in a fixed volume or as a part of a dynamically allocated bucket element,
- *node positions*—stored implicitly as a position in a fixed volume, explicitly as a set of coordinates, or even not stored.

An overview of sizes of the stored values is concentrated in Table 1 where for each algorithm the column headers represent the size of the structure in numbers of elements ($C, m$ or $n$)— the fixed size for static data structures or the maximal size for dynamic structures)—and the table entries represent the amount of data stored in each element of these structures.

For better understanding of the memory demands of the presented algorithms Tables 2, 3 and 5 show the modification-specific representation in amounts of bits/bytes in different levels of substitution and conversion, taking into account the last two lines of Table 1. Details are discussed in the explanations of the algorithm modifications (I–V) in Sections 4.2–4.6 and illustrated in Figures 1 to 5.

The proposed data structures can handle 16-bit datasets of maximal sizes up to $1024 \times 1024 \times 2048$ voxels. This upper limit is given by the storage of the temporary *label* and *position* in one 32-bit element. On the other hand, the real limits for the presented algorithms are the amount of RAM in a typical nowadays workstation and the largest possible dataset sizes presently available in the clinical praxis (about $512 \times 512 \times 1024$ voxels), i.e., the selected data structures do not form the restrictive part of the algorithm implementation.

### 4.2. Complete IFT with dequeuing and fixed volume for queue (I)

The complete IFT algorithm, as published by Lotufo and Falcao[1], is described in Section 3. To achieve the time complexity of $O(m+C)$ the priority queue with buckets is implemented as a double-linked list stored in the preallocated fixed volume of the same size equal to the size of the dataset (see Figure 1, where four voxels are stored in bucket 0).

This version uses all the operations described at the end of Section 3. Two arrays of pointers to both ends of the bucket-FIFOs are used for a constant time *DeQueueMin()* and *EnQueue(p, c)* while searching of the minimal element is done in *QueueNotEmpty()* in $O(C)$.

The double-linked list is necessary for a constant time *DeQueue(p)* for a node $p$ anywhere in the bucket queue; the complete list is stored in the volume of the same size as the size of the dataset to have a constant time for the *IsInQueue(p)* test and for a constant time location of the neighbors *GetNeighbor(v, dir)*. Each list element has also an entry with a current temporary cost $C(v)$ to achieve a constant time *GetCost(p)* (for details see also the first line in Tables 1–3). The only $O(C)$ time operation is the test *QueueNotEmpty()*, which sequentially searches the array of buckets for the first non-empty bucket. The complete search is performed once during the run of the algorithm.

| NUMBER OF ELEMENTS IN ARRAYS OF SIZE $C$, $m$ OR $n$ ALGORITHM | FIXED SIZE | | | | | DYNAMIC SIZE | | | | |
|---|---|---|---|---|---|---|---|---|---|---|
| | $C$-TEMP ptr | $n$-TEMP cost | $n$-TEMP flag | $n$-TEMP ptr | $n$-RESULT label | $n$-, $m$-TEMP cost | $n$-, $m$-TEMP label | $n$-, $m$-TEMP position | $n$-, $m$-TEMP ptr | $n$-, $m$-TEMP index |
| Complete IFT, deQing, fix volume (I) | 2 | 1 | 1 | 2 | 1 | – | – | – | – | – |
| Complete IFT, deQing, dynam. Q (II) | 2 | – | 1 | 1 | 1 | 1 | – | – | 2 | – |
| IFT no deQing, dynam. Q (III) | 2 | – | 1 | 1 | 1 | 1 | 1 | –* | 1 | – |
| IFT no max test, dynam. Q (IV) | 2 | – | 1 | – | 1 | – | (1) | 1 | 1 | – |
| IFT no max test, bricks (V) | 2 | – | 1 | – | 1 | – | (1) | 1 | $\frac{1**}{254}$ | $\frac{1**}{2*254}$ |
| Size of one element [bits] | 32 | 16 | 1 | 32 | 1 | 16 | (1) | 28–31 | 32 | 8–16 |
| Size of one element [Bytes] | 4 | 2 | $\frac{1}{8}$ | 4 | $\frac{1}{8}$ | 2 | 1 or 0 | 4 | 4 | 1–2 |

**Table 1:** *Memory necessary for the algorithms (in number of elements in each array) without an input dataset. The horizontal line separates the n- (upper part) and m- (lower part) size of the temporary volume. \*Position for (III) is computed indirectly from the values of pointers. \*\*Index and pointer sizes approximated for bricks of 256\*4 Bytes—one ptr and 2 indices for brick of 254 elements*

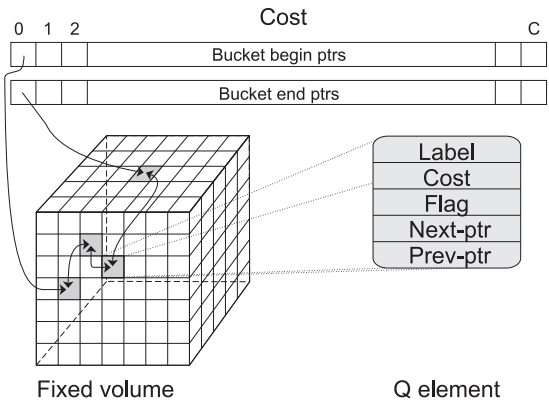

**Figure 1:** *Data structures for complete IFT algorithm with dequeuing and fixed volume for queue (I)*

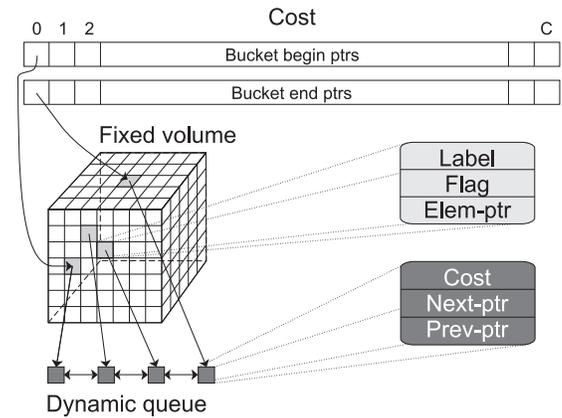

**Figure 2:** *Data structures for complete IFT algorithm with dequeuing and dynamical queue (II)*

### 4.3. Complete IFT with dequeuing and dynamical queue (II)

In an average case of the algorithm, the number of elements simultaneously stored in the queue is less than the place reserved in a fixed volume of $n$ entries (this feature is discussed in detail in Section 5.3). To reduce this extremely high fixed memory consumption a dynamical version of the priority queue was proposed.

The queue is now stored in a dynamically allocated double-linked list, each element of this list contains the value of the temporary cost $C(v)$ and two pointers for linking of the list (for details see the second line in Tables 1–3). As a direct access to the elements in the queue is still needed, a fixed volume of pointers to the elements in the list (queue) is still necessary. As no voxel is stored in the queue more than once the array reserved for the resulting labels can be "reused" also for the temporary labels the same way as in the complete algorithm (I). The set of operations and their complexities are equal to the fixed volume version from the previous Section 4.2.

### 4.4. IFT algorithm without dequeuing (III)

Lotufo and Falcao described also a possible simplification of the proposed algorithm. They propose to omit the dequeuing of the nodes with a higher temporary cost from the priority queue during the node-cost replacing operation (line 2cB of the complete IFT algorithm). This modification should simplify the queue data structure, as the operation *DeQueue(p)* does not need to be implemented.

However, herewith the uniqueness of the nodes in the queue is also lost, as the node (voxel) can be (and in many cases also is) stored more times than once in the queue. Therefore more different temporary costs and labels exist for one node at the same moment. It results in a repeated storage of the temporary cost $C(p)$ and temporary label in the queue within each queue entry. Multiple occurrences of the nodes in the queue are also common to both our modifications (described in Sections 4.5 and 4.6).

| Algorithm | Maximal memory consumption $O()$ in terms of $(C,n,m)$ | | |
| --- | --- | --- | --- |
| | Fixed temp | Fixed result | Dynamic temp |
| Complete IFT (I) | $(2C+2n)*ptr+n*(cost+flag)$ | $+n*label$ | $+0$ |
| Complete IFT (II) | $(2C+1n)*ptr+n*(flag)$ | $+n*label$ | $+n*(cost+2*ptr)$ |
| IFT without dequeue (III) | $(2C+1n)*ptr+n*(flag)$ | $+n*label$ | $+m*(cost+\underline{label}+ptr)$ |
| IFT without max (IV) | $(2C\phantom{+1n})*ptr+n*(flag)$ | $+n*label$ | $+m*(\underline{label+pos}+ptr)$ |
| IFT with bricks (V) | $(2C\phantom{+1n})*ptr+n*(flag)$ | $+n*label$ | $+m*(\underline{label+pos}+\frac{1}{254}*ptr+\frac{2}{254}*ind)$ |

**Table 2:** *Memory necessary for the algorithms in the worst case (in multiplicands of $(C,n,m)$ and data type pieces). Underlined label+pos are stored together, underlined label alone has to occupy a whole Byte. The fractional constant for ptr and ind is computed for the brick size of 254 elements*

| Algorithm | Maximal memory consumption $O()$ in terms of $(C,n,m)$ | | |
| --- | --- | --- | --- |
| | Fixed temp | Fixed result | Dynamic temp |
| Complete IFT (I) | $(2C+2n)*4+n*(2+\frac{1}{8})$ | $+n*\frac{1}{8}$ | $+0$ |
| Complete IFT (II) | $(2C+1n)*4+n*(\frac{1}{8})$ | $+n*\frac{1}{8}$ | $+n*(2+2*4)$ |
| IFT without dequeue (III) | $(2C+1n)*4+n*(\frac{1}{8})$ | $+n*\frac{1}{8}$ | $+m*(2+\underline{1}+4)$ |
| IFT without max (IV) | $(2C\phantom{+1n})*4+n*(\frac{1}{8})$ | $+n*\frac{1}{8}$ | $+m*(\underline{4}+4)$ |
| IFT with bricks (V) | $(2C\phantom{+1n})*4+n*(\frac{1}{8})$ | $+n*\frac{1}{8}$ | $+m*(\underline{4}+\frac{1}{254}*4+\frac{2}{254}*\underline{2})$ |

**Table 3:** *Table 2 after substitution of data type representation sizes according to the last line of Table 1. Underlined numbers arise from sharing of space (label+pos), allocation of the whole byte in the dynamical structures (label alone), or storage of indices in two Bytes*

But to preserve the constant time of the operation *GetNeighbor(v, dir)*, which is used in the comparison in the max test in line 2c of the IFT algorithm, a direct location of the neighbors in the queue is still necessary. This demand violates the presumption proposed by Lotufo and Falcao[1] that the *DeQueue(p)* operation can be completely removed (more precisely, this operation can be removed but the speed of the algorithm is simultaneously lost when the element in the queue is searched, or the volume of current minimal temporary costs is still necessary to be used, which needs memory and which still has to be continuously updated)!

The direct access of node neighbors can be achieved by handling of the volume of pointers to the elements in the queue; same as in the previous algorithm (II in Section 4.3). And the operation *DeQueue(p)* can be only simplified to move the pointer to the element with a lower value of $C(p)$ without physical removal of the previously pointed element (or to update the current minimal cost in case of a fixed volume of temporary costs).

To summarize the memory demands (see also the third line of Tables 1–3):

- Only one pointer in each queue element is now needed, as the elements are not removed from the queue and the queue is implemented as a single-linked list; but multiple entries of one node exist in the queue (double-entry example in Figure 4). Therefore, we store also the current cost and label in the queue but also multiple times with each node.
- In the fixed part of the data structure only one pointer in the fixed volume of pointers is saved (similar to (II)).

As a disadvantage the maximum queue length simultaneously increases from $O(n)$, as in modifications (I) and (II), to $O(m)$. For illustration of the data structures see Figure 3.

**4.5. IFT without dequeuing and without a max test (IV)**

To save even the memory of the fixed volume of pointers another modification of the IFT algorithm without dequeuing of elements is proposed in this paper where not only the dequeuing is omitted but also the maximum test (in line 2c). Simply, ALL nodes (their label and position) are stored according to the new values of the temporary path cost into the appropriate bucket in the priority queue where each bucket is implemented as a single-linked list. The cost is NOT stored in the queue—this information is intrinsic to each bucket, as it is the index in the bucket-begin- and bucket-end-pointer arrays (for details see Figure 4 and the fourth line of Table 1). As the FIFO returns (dequeues) the elements with lowest cost first and the algorithm

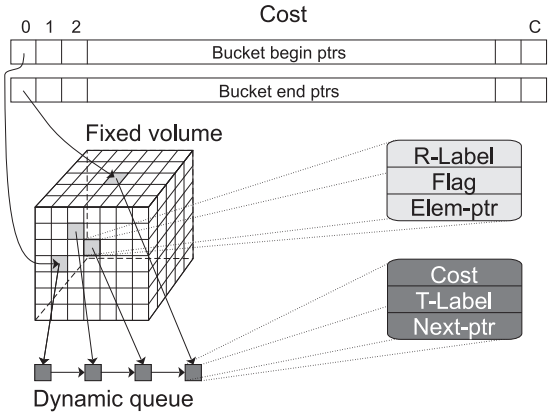

**Figure 3:** *Data structures for IFT algorithm without dequeuing (III)*

### 4.6. IFT without dequeuing and without a max test with bricks (V)

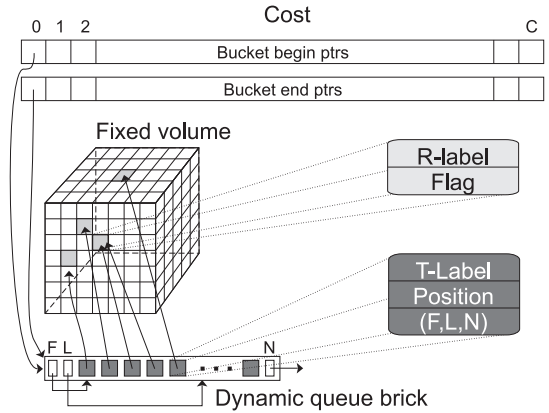

**Figure 5:** *Data structures for IFT algorithm without dequeuing and without max test with bricks (V)*

ignores the multiple entries after setting of the flag to DONE, the modified algorithm gives exactly the same results as the already discussed variants (I)–(III). Take note of a reverse direction of pointing from the queue elements into the volume (via position) and also a double-entry example in Figure 4. The volume now contains only the resulting label (R-label) and a flag, the temporary label and position are stored in the dynamically allocated queue elements.

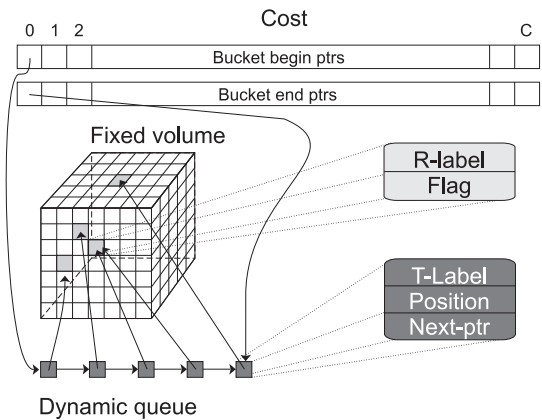

**Figure 4:** *Data structures for IFT algorithm with a dynamical queue without dequeuing and without max test (IV)*

As a disadvantage, we also can have a maximum of $O(m)$ elements in the queue. The exact number of elements in the queue is up to two times higher than in Section 4.4, as also the elements with a higher path cost will be stored somewhere in the end of the queue.

About one half of the memory is used for the pointers in the dynamical implementation of the FIFO via a list (for details see also the fourth line of Tables 2 or 3, the column Dynamic temp).

To save the memory used for pointers in the dynamical list in (IV), an allocation of memory in larger amounts was proposed – they are called *bricks* (see Figure 5, where the bucket "0" contains one brick with five entries, where two of them point to the same voxel). Each brick has one pointer to the next brick and two local indices for the first and the last (next free position behind the last) elements of the queue part stored in this brick (marked by F and L in Figure 5). 8-bit indices, allow the brick size (number of node entries stored in one brick) of max 256 elements. The common memory allocation scheme of multiples of 8 bytes (#pragma pack in VC++)[7] may prefer 254 entries, as 6-8 bytes are used for indices and a pointer to next brick (together $256 \times 8$ Bytes. For details see the fifth line of Tables 1–3, the column Dynamic temp).

To save the memory allocation times an empty brick management, which collects empty bricks in a separate "empty-brick" stack, is used.

### 5. Tests

Studies of computational and memory complexity exist for the shortest path forest problem, e.g., in general case by Ahuja et al.[4] and for the real road networks by Zhan[8]. The presented study is related to large medical volumetric datasets, where also other shortest path cost function is used (not sum, but maximum on the path, see Sections 2 and 5.2).

Five modifications of the IFT based watershed algorithm were discussed with segmentation of large 3D medical datasets in mind. This section discusses the estimations of asymptotic computational and memory complexities being made. The real case measurements on the 3D medical imaging datasets follow.

Properties of our typical volumetric medical datasets are characterized in Section 5.1. Memory and computational analysis of the worst and average case for each algorithm follow

| DATASET | | NUMBER OF | | ARC COSTS | | | ARC/NODE |
|---|---|---|---|---|---|---|---|
| NO | SIZE $X \times Y \times Z$ | NODES $n$ | ARCS $m$ | MAX | MEAN | SDEV | RATIO |
| 1 | $128 \times 128 \times 79$ | 1 294 336 | 3 846 400 | 4 095 | 50.40 | 152.05 | 2.972 |
| 2 | $256 \times 256 \times 48$ | 3 145 728 | 9 347 072 | 568 | 9.98 | 20.82 | 2.971 |
| 3 | $255 \times 255 \times 81$ | 5 267 025 | 15 694 740 | 3 921 | 26.18 | 68.46 | 2.980 |
| 4 | $256 \times 256 \times 444$ | 28 097 984 | 87 001 088 | 1 627 | 23.81 | 78.60 | 2.990 |
| 5 | $512 \times 512 \times 125$ | 32 768 000 | 97 913 856 | 4 095 | 24.27 | 80.57 | 2,988 |
| 6 | $512 \times 512 \times 169$ | 44 302 336 | 132 471 808 | 1 483 | 23.00 | 50.25 | 2.990 |

**Table 4:** *Characteristics of medical datasets used in the study*

on in Sections 5.2 and 5.3. Results of tests are summarized in Section 5.4.

### 5.1. Datasets

The type of the dataset (its size and precision of stored values) is important for the algorithm implementation, as it determines the memory demands for temporary data structures, e.g., the representation of stored costs and data structures for the priority queue, as discussed in Section 4.

The implementation of IFT algorithm has been tested and tuned on 3D datasets (volumes) acquired by computer tomography (CT) and magnetic resonance imaging (MRI). The size of these data volumes is $r \times r \times S$, where inter-slice resolution $r$ varies from 128 to 512 and the number of slices $S$ varies from 48 to 444. Characteristics of medical datasets used for tests such as maximal, mean and standard deviation of the arcs are listed in Table 4. The datasets are ordered according to the number of nodes. Example images of three of them are shown in Figure 6.

The practical maximal size of dataset the algorithm can handle is given by the operation memory of PC's at the present time limited to about 1GB, i.e., about 512–680 slices (512 to omit swapping, which occurs if about 75% memory is used). Nevertheless, the proposed data structures allow to use datasets up to $1024 \times 1024 \times 2048$ slices, as they use one bit for the label and 31 bits are left for the position. Higher dataset sizes or more than one label value cause an increase in memory complexity, as the label has to allocate a separate byte.

As already noted in Section 4.1, precision $C_p$ of the 12–16-bit datasets gives an upper bound for the maximal difference $C$. That means a fixed size of a bucket pointer array of 4 096–65 536 pairs of bucket pointers can be used if the maximal difference of values in the dataset were not directly available and had to be computed (in an additional computational cost $O(m)$). This roughly approximated array of bucket pointers occupies about 0.5MB of memory, nevertheless this amount of memory in comparison to the whole memory demands of the algorithm is in practical terms inconsiderable.

Zhan[8] draws the attention to different types of graphs for different tasks, classified according to *arc-to-node ratio*, as this parameter directly implies the complexity of the algorithm. Zhan[8] also warns that the measurements of the performance of the shortest-path algorithms have been done on artificial networks with arc-to-node ratio up to ten, where the real road network graphs achieve the value of only approximately three.

The *arc-to-node ratio* ($m/n$) is almost equal to three for a volumetric graph, because most of the nodes belong to the inner voxels with six neighbors and each arc is processed only once. Arc-to-node ratio of medical datasets used for tests is listed in the last column Table 4.

For real road networks, one of the fastest algorithms was the Dijkstra's algorithm implemented with appropriate buckets. As volumetric datasets have an equal arc-to-node ratio, it can be deduced that the concentration of the work on a good implementation of the Dijkstra's algorithm with buckets was a good choice.

### 5.2. The worst case analysis

All presented modifications of the algorithm have the same $O(m+C)$ *time complexity*, they differ in the constant factor in the average and maximal running times and in the maximal memory complexity.

As stated in Section 2, the path cost is for the watershed algorithm computed by the IFT defined as the *maximal arc weight* in the path. The "classical" shortest path forest algorithms (discussed by Ahuja[4] or Zhan[8]) use a different definition of the path cost as the *sum* of the arc weights in the path. The "classical" variants therefore need $n$ runs through the queue in the worst case and reach time complexity of $O(m+nf(C))$ – with the multiplicative term $n$. The IFT-WS by Lotufo et al.[1] needs only one complete run and the time complexity is simplified to $O(m+C)$.

The worst case *memory complexity* of presented algorithm modifications varies from $O(n)$ to $O(m)$:

- The algorithm variants *with dequeuing* (the complete IFT with a fixed volume (I) and the complete IFT with dynamical queue (II)) have a fixed size of the supporting data structures of $O(n+C)$ and a fixed or maximal size for the queue $O(n)$. Therefore is the memory consumption of $O(n+C)$.
- The variants *without dequeuing* (IFT without dequeuing (III), IFT without dequeuing without the max test (IV) and IFT without dequeuing without the max test and with bricks (V)) have all the fixed size of the supporting data structures of $O(n+C)$ and, as the elements are not dequeued, the maximal size for the queue of $O(m)$. Therefore is the memory consumption of $O(m+n+C) = O(m+C)$.

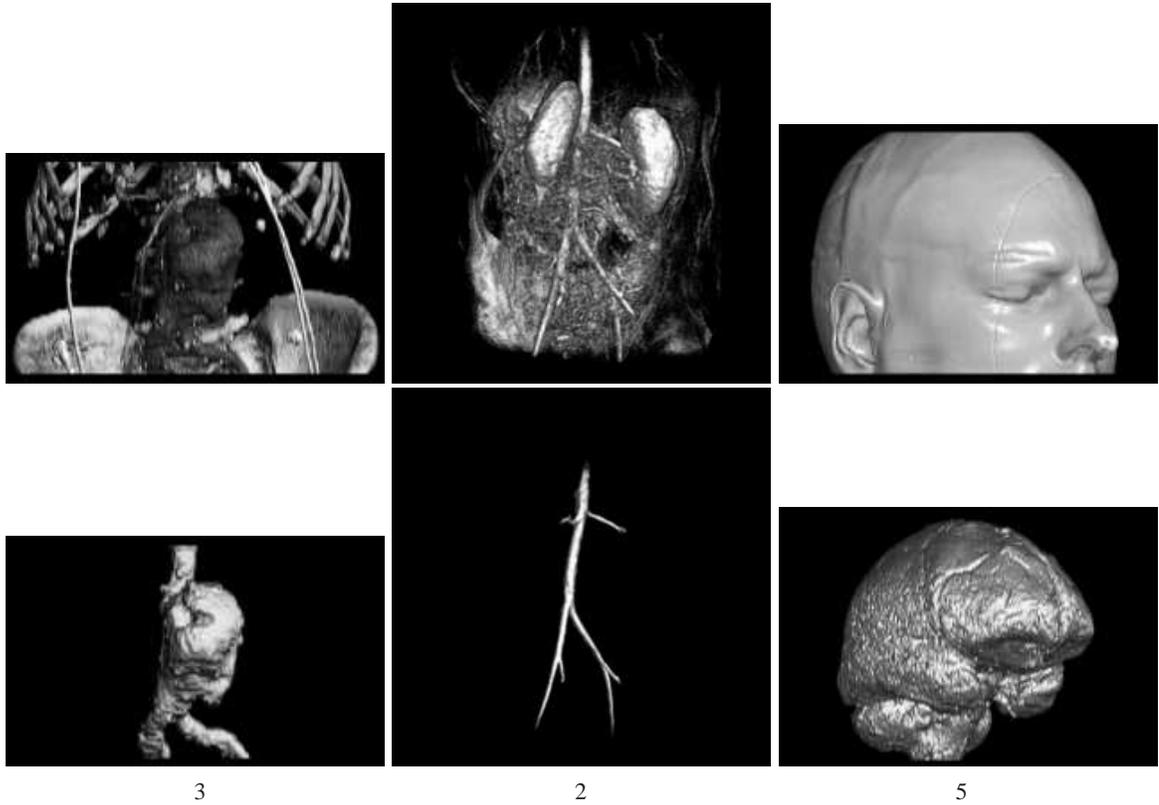

| 3 | 2 | 5 |

**Figure 6:** *Examples of the datasets used in tests. Top row shows a windowed view of input datasets, bottom row segmented structures. Numbers are according to Table 4. (Datasets with courtesy of TIANI Medgraph GmbH, Austria)*

Table 5 shows the maximal (worst case) memory sizes of five tested algorithms (I–V) in multiples of $C, n$ and $m$ after all substitutions and conversions. These maximal sizes are reached only in a special case of input markers set in a chessboard pattern, which cause planning of all arcs simultaneously into the queue. In real situations, the dynamical part depends on the real queue length $m'$ or $n'$, which is smaller than $m$ or $n$.

Before testing the algorithms, a comparison of the maximal number of elements the algorithms can handle in the same amount of memory as the complete IFT algorithm (I) was performed (see the last column of Table 5). For instance, the variant with bricks (V) can handle up to 82% of the elements stored in the queue (of length about $3n$). The comparison is based on the premise that the queues will be filled similarly in all variants (I–V), which must not be true, as the storage strategies differ.

It is possible to see, that Lotufo and Falcao's[1] idea for the data structures simplification by omitting the dequeuing is for medical datasets not so good (more precisely, a better way of its implementation is not known to the authors of this paper). Omitting the *DeQueue(p)* operation with survival of the maximum test allow only 29% of arcs stored in the queue to be han-dled in the same amount of memory as (I) (Table 5). Or, to save memory, the volume of pointers used for the direct access of the queue elements in the max test can be also omitted but it would slow down the algorithm.

| | MAX MEMORY $O() = f(C,n,m)$ | | | RELATIVE |
|---|---|---|---|---|
| ALG | FIXED PART FOR ALL | DIFF | DYNAMICAL PART | Q LEN [%] |
| I | $8C + \frac{1}{4}n$ | 10n | — | 100% |
| II | $8C + \frac{1}{4}n$ | 4n | $10n$ | 60% |
| III | $8C + \frac{1}{4}n$ | 4n | $7m \approx 21n$ | 29% |
| IV | $8C + \frac{1}{4}n$ | — | $8m \approx 24n$ | 42% |
| V | $8C + \frac{1}{4}n$ | — | $4.0314m \approx 12.1n$ | 82% |

**Table 5:** *Table 3 after conversion in multiplicands of bytes. The last column shows the maximal amount of elements that can be processed in the same memory as used by the complete IFT algorithm with the queue in the fixed volume (I). (Algorithm (I, II) in percents of $n$, (III–V) in percents of $m \approx 3n$)*

| DS NO | ALG NO | FIX TEMP | | | DYNAM | DYNAMIC USED BY ELEMENTS | | | SUM MEMORY | | | MEM |
| --- | --- | --- | --- | --- | --- | --- | --- | --- | --- | --- | --- | --- |
| | | 8C [MB] | 1/4n [MB] | VOL [MB] | MAX [MB] | NO ELEMS ($m'$ OR $n'$) | FILLING [% OF Q] | MEM [MB] | MAX [MB] | USED [MB] | [% MAX] | OF I [%] |
| 1 | I | 0.5 | 0.3 | 12.34 | – | 181 243 | **14.0%** | – | 13.2 | 13.2 | 100% | **100%** |
| | II | 0.5 | 0.3 | 4.94 | 12.3 | 181 243 | **14.0%** | 1.7 | 18.1 | 7.5 | 41% | **57%** |
| | III | 0.5 | 0.3 | 4.94 | 25.9 | 558 862 | **14.5%** | 3.7 | 31.7 | 9.5 | 30% | **72%** |
| | IV | 0.5 | 0.3 | – | 29.6 | 1 029 308 | **26.8%** | 7.9 | 30.4 | 8.7 | 28% | **66%** |
| | V | 0.5 | 0.3 | – | 14.9 | 1 029 308 | **26.8%** | 4.9 | 15.7 | 5.6 | 30% | **42%** |
| 2 | I | 0.5 | 0.8 | 30.00 | – | 1 073 959 | **34.1%** | – | 31.3 | 31.3 | 100% | **100%** |
| | II | 0.5 | 0.8 | 12.00 | 30.0 | 1 073 959 | **34.1%** | 10.2 | 43.3 | 23.5 | 54% | **75%** |
| | III | 0.5 | 0.8 | 12.00 | 63.0 | 1 562 900 | **16.7%** | 10.4 | 76.3 | 23.7 | 31% | **76%** |
| | IV | 0.5 | 0.8 | – | 72.0 | 2 503 049 | **26.8%** | 19.1 | 73.3 | 20.3 | 28% | **65%** |
| | V | 0.5 | 0.8 | – | 36.3 | 2 503 049 | **26.8%** | 9.9 | 37.6 | 11.1 | 29% | **36%** |
| 3 | I | 0.5 | 1.3 | 50.23 | – | 757 453 | **14.4%** | – | 52.0 | 52.0 | 100% | **100%** |
| | II | 0.5 | 1.3 | 20.09 | 50.2 | 757 453 | **14.4%** | 7.2 | 72.1 | 29.1 | 40% | **56%** |
| | III | 0.5 | 1.3 | 20.09 | 105.5 | 2 629 578 | **16.8%** | 17.6 | 127.3 | 39.4 | 31% | **76%** |
| | IV | 0.5 | 1.3 | – | 120.6 | 4 641 552 | **29.6%** | 35.4 | 122.3 | 37.2 | 30% | **71%** |
| | V | 0.5 | 1.3 | – | 60.8 | 4 641 552 | **29.6%** | 18 7 | 62.5 | 20.5 | 31% | **40%** |
| 4 | I | 0.5 | 6.9 | 277.50 | – | – | – | – | 284.9 | 284.9 | 100% | **100%** |
| | V | 0.5 | 6.9 | – | 335.8 | 11 735 946 | **13.5%** | 45.4 | 343.2 | 52.8 | 15% | **19%** |
| 5 | I | 0.5 | 7.8 | 312.50 | – | – | – | – | 320.8 | 320.8 | 100% | **100%** |
| | V | 0.5 | 7.8 | – | 378.1 | 26 530 954 | **27.1%** | 102.7 | 386.4 | 111.0 | 29% | **35%** |
| 6 | I | 0.5 | 10.6 | 422.50 | – | – | – | – | 433.6 | 433.6 | 100% | **100%** |
| | V | 0.5 | 10.6 | – | 511.2 | 47 513 276 | **35.9%** | 181 9 | 522.3 | 193.0 | 37% | **45%** |

**Table 6:** *Measurements of memory consumption for 3D medical datasets. Values for measurements on datasets number 4–6 that did not fit into memory of our implementation were omitted (the queue length for algorithm (I) and complete lines for (II–IV))*

### 5.3. Average case complexity analysis for medical datasets

Six volumetric medical datasets were tested and measured. The tests were designed to verify the derived time and memory complexities, the average filling of the queue and, for the last modification, also the average brick fill to check the suitability of the selected brick size. In all cases, $C = MaxDiff$ was used, as this value was directly available for each dataset.

At first a test-bed implementation of all algorithms was done by means of general container library[7]. The tests showed that the memory consumption of algorithms (I–IV) was too high for the 3D medical datasets and also that the dynamical space management consumes too much processor power (up to 10-times slower in our test implementation). After these preliminary tests, the last variant (V) was carefully implemented with the specifically-designed data structures.

Therefore the computational times were measured only for the last implementation of the IFT algorithm (V) and they are collected for the presented 3D medical datasets in the last column of Table 7. The linear time complexity is from these numbers obvious.

The memory consumption was measured for all variants (I–V). As mentioned in Section 5.2, the extreme situation (maximum memory consumption) when the maximum of $m$ elements simultaneously stored in the queue were reached is practically impossible for real cases of the volumetric datasets with manually set markers.

In the real situation, when the user puts markers in the 3D volume and starts the algorithm, the memory consumption of all the modifications (II–V) is less than the needs of complete IFT (I). The first bold column in Table 6 shows the relative queue fill reached in our tests, which varies from 14 to 34% for variants (I, II) to 26 to 36% for variant (V). A comparison of these values with the last column of Table 5 shows that the last variant (V) can handle all the cases with a large reserve in the same amount of memory as (I) but the variant (III) can not.

More interesting is the comparison of memory usage for the algorithm modifications, concentrated in the second bold column in Table 6. It is possible to see, e.g., that the algorithm variant that stores all arcs in the queue and performs no comparisons of previously stored costs (V) needs only about 19 to 45% of the memory of the reference complete implementation with a fixed volume (I). This is the best implemented modification.

The amount of elements in the queue and therefore the whole memory depend substantially on the segmented structure, given markers and the character of the dataset. It seems that small homogeneous structures in a homogeneous background need a lower amount of elements simultaneously in the queue (Dataset No. 4 with 13.5% of elements) but up to now the exact correlation between these parameters and the average queue fill have not been found. This is a topic for a further research.

The last test was designed to find the average brick fill for the bricks with the selected size of 254 elements. For the version with bricks (V), the largest amount of bricks is used in such a case when all buckets contain a brick with one element and the rest is more or less full. But this situation changes the memory demands in amount of ones of percent for large datasets and therefore can be omitted in the tests. Results of the tests are collected in Table 7. The brick fill is typically between $\frac{4}{5}$ and 1, which means that the brick size is relatively optimal and the size of brick was chosen correctly.

| DS No | NUMBER OF ELEMS IN Q | BRICKS | AVERAGE BRICK SIZE | COMPUT TIME |
|---|---|---|---|---|
| 1 | 1 029 308 | 4 970 | 205.8 | 2.3 |
| 2 | 2 503 049 | 10 126 | 245.5 | 6.0 |
| 3 | 4 641 552 | 19 174 | 242.1 | 10.1 |
| 4 | 11 735 946 | 46 502 | 252.4 | 51.6 |
| 5 | 26 530 954 | 105 184 | 252.2 | 67.8 |
| 6 | 47 513 276 | 186 232 | 253.1 | 108.2 |

**Table 7:** *Tests of the brick filling for all datasets in the study (Continuing of Table 4 for algorithm (V)). Number of bricks is herewith equal to amount of memory used for bricks in [kB]. The last column shows the computatonal time for (V) in [s]*

Final note: To be exact it has to be mentioned that the memory used by a compiler for handling of the dynamical memory (in the compiler documentation[7] called the "maintenance cost") was not taken into consideration.

### 5.4. Summary of tests results

The complete IFT algorithm implemented with a fixed volume (I) served as a reference for comparison of the algorithms. It uses a fixed amount of memory, which is given by number of voxels in the dataset.

The dynamical version of the complete algorithm (II) uses about 56–75% of the memory of (I). In spite of the fact that this algorithm replaces the elements in the queue and therefore just stores each node in the queue at most once, the saved amount of memory is not substantial. Handling of the dynamical variables results in slower computational times (up to 10-times slower in our test-bed implementation).

By omitting of the node replacing (line 2cB/C), as proposed Lotufo and Falcao[1] (III), the implementation saves less memory then (II), as omitting of dequeuing causes multiple node entries in the queue and these entries need a relatively large amount of memory for storage. This variant needs in an average case about 72–76% of the memory of (I).

Not much better is the first max-test simplification (IV), which saves the fixed part of the temporary memory (pointers to the queue elements) by omitting of the comparison in the line 2c. The saved memory is nearly all used by the multiple entries of nodes with a higher cost. The average memory consumption is about 65–71% of the memory of (I).

Significantly (two times) better in the worst case and even the best implementation in average case for real 3D medical datasets is the second max-test simplification with bricks (V). It also saves the fixed part of the temporary memory but, by better memory allocation in larger amounts, it can handle up to 80% of the maximal queue length in the same amount of the memory of (I) and in an average case needs only 19–45% of the memory of (I).

### 6. Conclusions and future work

The paper presents five variations of the Image Foresting Transform (IFT) algorithm implemented with the aim of performing segmentation by watershed from markers for large 3D medical datasets in a limited amount of operational memory.

An optimization of the memory consumption was performed and the best algorithm variant needs in an average case about 19–45% of memory of the complete IFT algorithm implemented according to Lotufo and Falcao.

The best implementation variant allows segmentation of datasets up to sizes of $1024 \times 1024 \times 2048$ voxels but the memory of currently available medical workstations based on PC technology allows a segmentation of significantly smaller datasets of sizes up to $512 \times 512 \times 512$–680 voxels.

An interesting area for the future work is finding correlations between dataset parameters (such as that in Table 4) and the number and the character of input markers to the maximal queue length. If this correlation exists, it can be used for prediction of the overall memory consumption of the segmentation algorithm.

A modification of arc-weights in an interslice direction $z$ can also be tested on the basis of the presumption that if the sampling distance in $z$-direction is higher, then also the absolute difference is higher and therefore the flow in the 2D slice may be preferred over jumping to the neighboring slice.

Another interesting area is an application of hierarchical techniques for speeding up the watershed algorithm itself and the overall user interaction. At present, a combination of 2D and 3D watersheds has been tested, where the complete results of 2D watersheds have been used as markers for 3D watershed. This combination has given promising results.

The presented algorithm should be also further validated in a clinical praxis.

**Acknowledgments**

This work has been done in the VRVis research center, Vienna, Austria (*http://www.vrvis.at*), which is partly funded by the Austrian government research program Kplus. The Datasets used for tests in this paper are courtesy of TIANI Medgraph GmbH, Austria, *http://www.tiani.com*.